\documentclass[aps,pra,reprint,amsmath,amssymb,nofootinbib,floatfix]{revtex4-2}
\usepackage[debug=false,lineno=false]{fixes}        % Produces PGFplots
\usepackage{physics}
\usepackage{enumitem}
\usepackage{graphicx}	        % Include figure files
\usepackage[table]{xcolor}            % Colours
\definecolor{PRDblue}{HTML}{2E3092}
\usepackage[colorlinks=true, allcolors=PRDblue]{hyperref} 
\usepackage{cleveref}
\usepackage{orcidlink}
\usepackage{fontawesome}

\providecommand{\ii}{\mathrm{i}}
\providecommand{\ee}{\mathrm{e}}
\providecommand{\vc}{\vb*}
\providecommand{\uv}{\vu*}

\newcommand{\bessel}[1]{\operatorname{J}_{#1}}
\newcommand{\neumann}[1]{\operatorname{Y}_{#1}}
\newcommand{\macdonald}[1]{\operatorname{K}_{#1}}
\newcommand{\hankel}[2][(1)]{\operatorname{H}^{#1}_{#2}}
\newcommand{\harmol}[1]{\operatorname{Z}_{#1}}

\begin{document}
\title{A robust and efficient method to calculate electromagnetic modes on a cylindrical step-index nanofibre}
\author{Sebastian \surname{Golat}\,\orcidlink{0000-0003-3947-7634}}
\email{sebastian.1.golat@kcl.ac.uk}

\author{Francisco J. \surname{Rodr\'iguez-Fortu\~no}\,\orcidlink{0000-0002-4555-1186}}
\email{francisco.rodriguez\_fortuno@kcl.ac.uk}
\affiliation{
Department of Physics and London Centre for Nanotechnology, 
King's College London, Strand, London WC2R 2LS, UK}

\date{\today}
\begin{abstract}

The accurate calculation of guided electromagnetic modes in optical nanofibres is critical for applications in nanophotonics, from quantum interfaces to vectorial light sensing. Standard textbook methods rely on solving a $4\times4$ matrix eigenvalue problem to find the modal fields. While widely used, this approach has a subtle but significant flaw: the final determination of the field amplitudes requires finding the numerical null space of a theoretically singular matrix, an ill-conditioned problem that introduces large relative errors in the small but physically crucial longitudinal field components.
In this work, we introduce a fundamentally more robust and efficient semi-analytical method. By starting from the foundational symmetries of the cylindrical waveguide and employing a judicious normalisation of the field amplitudes, we demonstrate that the problem can be analytically reduced to a much simpler $2\times2$ system. This reformulation yields two decisive advantages: the dispersion relation is obtained numerically from a simple and well-behaved transcendental equation, and more importantly, the modal field amplitudes are subsequently determined \emph{analytically}. Our approach completely bypasses the numerical null space calculation, thereby ensuring the accuracy of the full vectorial field structure. This method provides a powerful and reliable tool for the design and analysis of nanofibre-based devices, particularly for applications in chiral quantum optics and nanophotonics where precise knowledge of field polarisation and specifically of the longitudinal components is paramount.

\end{abstract}

\maketitle%
\section{Introduction}%
Nanofibres are ultrathin cylindrical waveguides whose diameters are comparable to, or smaller than, the wavelength of the guided light.  Because a substantial fraction of every guided mode resides in the evanescent field outside the glass, nanofibres serve as versatile nanophotonic interfaces.  They have enabled optical trapping, propulsion, and rotation of colloids, bacteria, and other microparticles \cite{Kamath2023}, ultra-sensitive surface and volume biosensing of molecules and quantum dots \cite{Morrissey2013}, and the stable trapping and interrogation of laser-cooled atoms only a few hundred nanometres from the fibre surface—crucial for quantum memories and chiral quantum-optics experiments \cite{Vetsch2010,Reitz2013}. Nanofibre sensors are industry-proven \cite{Zhang2020,Luo2021}, known for superior performance, low sample volumes, and strong analyte interactions---driving the fastest-growing segment of the optical-sensor market \cite{MordorIntelligence2025}.

All of these applications require an accurate description of the electromagnetic modes supported by the nanofibre.  The conventional approach starts by solving Maxwell’s equations in the core and the cladding and then enforcing continuity of the tangential fields $E_{z},H_{z},E_{\phi},H_{\phi}$ at the boundary.  Writing the fields (electric and magnetic fields, both in the core and cladding regions) in terms of four constants $(A,B,C,D)$ multiplied by Bessel or modified Bessel functions leads to a homogeneous linear system of four equations.  This is a standard procedure described in both traditional textbooks and recent research papers \cite{Balanis1989,Picardi2018}. Setting the determinant of the resulting $4\times4$ matrix to zero yields the dispersion relation for the propagation constant $k_z$; a second numerical step then finds the null space to determine $(A,B,C,D)$ and hence the modal field distributions and polarisation.

Although it is standard, this two-step procedure suffers from numerical instabilities: the longitudinal field components---responsible for chiral and other vectorial light–matter effects---are much smaller than the transverse fields, so small numerical errors in the null space translate into large relative errors in the quantities of interest.  In this work, we realised that, with a convenient normalisation of the field amplitudes, the textbook $4\times4$ system can be shown to be equivalent to:
\begin{equation}\label{eq:four-by-four}
\begin{pmatrix}
1 & 0 & -1 & 0 
\\ 0 & 1 & 0 & -1
\\ \frac{\ell n_\text{e} }{u^2} & \frac{\ii {\mu_1}}{u}\frac{\bessel{\ell}'}{\bessel{\ell}}  & \frac{\ell n_\text{e} }{w^2} & \frac{\ii {\mu_2}}{w}\frac{\macdonald{\ell}'}{\macdonald{\ell}}   
\\  { \frac{{\varepsilon_1}}{\ii u}\frac{\bessel{\ell}'}{\bessel{\ell}}}  & \frac{\ell n_\text{e} }{u^2}& \frac{{\varepsilon_2}}{\ii w} \frac{\macdonald{\ell}'}{\macdonald{\ell}} & \frac{\ell n_\text{e} }{w^2}
\end{pmatrix}
\!\!
\begin{pmatrix}
A \\ B \\ C \\ D
\end{pmatrix}
\! = \!
\begin{pmatrix}
0 \\ 0 \\ 0 \\ 0
\end{pmatrix},
\end{equation}
whose first two rows immediately give $A=C$ and $B=D$.  All the remaining symbols will be thoroughly defined later. The physical problem therefore effectively reduces to a $2\times2$ system that we will concisely derive from first principles.  This reformulation, being the main novelty of this work, offers two decisive advantages:

\begin{enumerate}
\item The dispersion relation is obtained by equating the determinant of a $2\times2$ matrix to zero, greatly simplifying numerical root-finding.
\item Once the propagation constant $k_z$ and frequency $\omega$ are known, the field amplitudes $A$ and $B$ follow \emph{analytically} from the remaining two equations---thereby eliminating the ill-conditioned null space calculation and preserving the accuracy of the longitudinal components.
\end{enumerate}

The remainder of the paper derives this reduced system from first principles, starting from symmetry arguments alone, benchmarks its accuracy, and illustrates its benefits.

The method described here is implemented in our open-source Python package \href{https://github.com/Sevastienn/anafibre}{\textcolor{black}{\texttt{Anafibre} \faGithub}} \cite{anafibre}, which provides routines for solving the dispersion relation, evaluating full vectorial fields, and normalising modes. 

\section{Cylindrical symmetry and mode expansion}
The method to find the field solutions of a nanofibre is well-known, but in order to arrive at a robust, simplified method, it is worth re-deriving the procedure from first principles. We will consider a step-index cylindrical waveguide, described by relative permittivity and permeability that abruptly change at the boundary. If $\rho$ is the radial coordinate, then:
\begin{equation*}
    \alpha_\text{r}=\left\{\begin{array}{ll}
\alpha_1, & \text { for } \rho<\rho_0 \\
\alpha_2, & \text { for } \rho>\rho_0
\end{array}\qq{where}\alpha_\text{r}\in\{\varepsilon_\text{r},\mu_\text{r}\}\right..
\end{equation*}
% \paco{[See later comment in \cref{eq:phi}, I think $\alpha_{\mathrm{r}1}$ and $\alpha_{\mathrm{r}2}$ would be clearer]}
In this work, we restrict attention to monochromatic fields, which ensures time-translational symmetry of the problem, but the generalisation is straightforward. Time translations are transformations generated by the Hamiltonian energy operator $\hat{H}=\ii\hbar\partial_t$.
This waveguide exhibits cylindrical symmetry characterised by invariance under translations along the fibre axis ($z$-axis) and rotations around it. Translations along the $z$-axis are generated by the linear momentum operator $\hat{p}_z = -\ii \hbar \partial_z$, while rotations about this axis are generated by the angular momentum operator $\hat{J}_z = \hat{L}_z + \hat{S}_z$, with orbital angular momentum $\hat{L}_z = -\ii \hbar \partial_\varphi$. The spin angular momentum operator $\hat{S}_z = \ii \hbar\, \uv{z}\times$ accounts for rotations of the field itself, provided the field is vectorial rather than scalar. Since both translations and rotations correspond to symmetries of the waveguide, the fields can be most conveniently expressed using a mode expansion into eigenvectors $\vc{F}_{\ell m}$ (either $\vc{E}_{\ell m}$ or $\vc{H}_{\ell m}$) that simultaneously diagonalise these symmetry operators. 
This means that these eigenvectors are separable functions $f(t,\rho,\varphi,z)=T(t)R(\rho)\Phi(\varphi)Z(z)$: 
\begin{equation}\label{eq:ansatz}
    \vc{F}_{\ell m}(t,\vc{r})=\ee^{-\ii \omega t}\sum_{s=-1}^1 F_{\ell m}^{(s)}(\rho)\,\uv{e}_{s}\;\ee^{\ii (\ell-s)\varphi}\ee^{\ii k_z z}
    \,,
\end{equation}
where $\uv{e}_0=\uv{z}$ and $\uv{e}_{\pm1}=(\uv{x}\pm\ii\uv{y})/\sqrt{2}$ are eigenvectors of the spin operator $\hat{S}_z\uv{e}_s=\hbar s\uv{e}_s$, $\ee^{\ii (\ell-s)\varphi}$ are eigenvectors of the orbital angular momentum, such that:
\begin{equation}
    \hat{J}_z\vc{F}_{\ell m}={(\hat{L}_z+\hat{S}_z)}\vc{F}_{\ell m}=\hbar \ell\vc{F}_{\ell m}\,,
\end{equation}
where $\ell\in\{0,1,2,\ldots\}$ is the azimuthal order, and $\ee^{\ii k_z z}$ is an eigenvector of the linear momentum:
\begin{equation}
    \hat{p}_z\vc{F}_{\ell m}=-\ii\hbar\partial_z\vc{F}_{\ell m}=\hbar (k_z)_{\ell m}\vc{F}_{\ell m}\,,
\end{equation}
where the propagation constant $(k_z)_{\ell m}$ has to be determined from the dispersion relation, which for each $\ell$ has more than one solution labelled by $m\in\{1,2,3,\ldots\}$. Finally the $\ee^{-\ii \omega t}$ is an eigenvector of energy:
\begin{equation}
    \hat{H}\vc{F}_{\ell m}=\ii\hbar\partial_t\vc{F}_{\ell m}=\hbar \omega\vc{F}_{\ell m}\,.
\end{equation}
The physical fields are then $\vc{\mathcal{F}}_{\ell m}(t,\vc{r})=\Re\qty[ \vc{F}_{\ell m}(t,\vc{r})]$ (either $\vc{\mathcal{E}}_{\ell m}$ or $\vc{\mathcal{H}}_{\ell m}$).
By plugging the ansatz \cref{eq:ansatz} into the Helmholtz equation, we can easily verify that the  scalar radial dependencies, $F_{\ell m}^{(s)}(\rho)$ (either $E_{\ell m}^{(s)}$, $H_{\ell m}^{(s)}$), satisfy a Bessel equation:
\begin{equation}
    \frac{1}{\rho}{\dv{\rho}}\qty(\rho\dv{F_{\ell m}^{(s)}}{\rho})+\qty[\kappa_{1,2}^2+\frac{(\ell-s)^2}{\rho^2}]F_{\ell m}^{(s)}=0\,,
\end{equation}
where $\kappa^2_{1,2}=k_{1,2}^2-k_z^2$ is the radial wavenumber defined for each medium via the wavenumber $k_{1,2}=n_{1,2} k_0$, with the refractive index of each medium $n_{1,2}=\sqrt{\varepsilon_{1,2} \mu_{1,2}}$ and the wavenumber of free space $k_0=\omega/c_0=2 \pi/\lambda_0$. If $\kappa_{1,2}\neq0$, this equation is solved by the linear combination of Bessel functions $\bessel{\ell-s}(\kappa_{1,2}\rho)$ and $\neumann{\ell-s}(\kappa_{1,2}\rho)$, or of Hankel functions $\hankel[(1)]{\ell-s}(\kappa_{1,2}\rho)$ and $\hankel[(2)]{\ell-s}(\kappa_{1,2}\rho)$.
For a lossless waveguide, one would expect the dependence in the core of the fibre to be regular and oscillatory ($\kappa_1^2>0$), while the dependence in the cladding would be decaying ($\kappa_2^2<0$), but the equations admit arbitrary complex values in $\kappa_1$ and $\kappa_2$
\begin{equation}\label{eq:radial}
    F_{\ell m}^{(s)}(\rho)=
    \begin{cases}
        \Tilde{F}_{\ell m}^{(s)}\;\dfrac{\bessel{\ell-s}(\kappa_1\rho)}{\bessel{\ell}(\kappa_1\rho_0)}& \text{if} \quad\rho<\rho_0
        \\\Tilde{F}_{\ell m}^{(s)}\;\dfrac{\hankel{\ell-s}(\kappa_2\rho)}{\hankel{\ell}(\kappa_2\rho_0)}& \text{if} \quad\rho>\rho_0
    \end{cases},
\end{equation}
where $\tilde{F}_{\ell m}^{(s)}$ (either $\tilde{E}_{\ell m}^{(s)}$, $\tilde{H}_{\ell m}^{(s)}$) are scalar amplitudes.
Notice that we normalise our radial dependence by the Bessel/Hankel function at the boundary with $s=0$.\footnote{One can also write this in terms of the Macdonald function $\macdonald{\ell}$ using the relationship valid for \(\Re{(z)}>0\):
\[
\macdonald{\ell}(z)=\frac{\pi}{2}\ii^{\ell+1}\hankel{\ell}(\ii z)\,,
\]
like in \cite{Balanis1989}. However, Hankel functions lead to a simpler form of \cref{eq:spin_amplitudes} and can be more intuitive in the case of lossy media.} This has the crucial advantage that we can identify the longitudinal components {($s=0$)} of the fields at the boundary with \emph{two} amplitudes (rather than \emph{four}):
\begin{equation}\label{eq:z_component}
    E_{\ell m}^{(0)}(\rho_0)\mathop{=}\tilde{E}_{\ell m}^{(0)}=\frac{A_{\ell m}}{\sqrt{\varepsilon_0}}\,,\quad
        H_{\ell m}^{(0)}(\rho_0)\mathop{=}\tilde{H}_{\ell m}^{(0)}=\frac{B_{\ell m}}{\sqrt{\mu_0}}\,.
\end{equation}
The factors of $\sqrt{\varepsilon_0}$ and $\sqrt{\mu_0}$ ensure that $|A_{\ell m}|^2$ and $|B_{\ell m}|^2$ have units of energy density.
The remaining components with $s=\pm1$ can be found in terms of these amplitudes by plugging our ansatz into Maxwell's equations which leads to (see \cref{app:max}):
\begin{equation}\label{eq:spin_amplitudes}
    \begin{aligned}
    \Tilde{E}_{\ell m}^{(\pm1)}&=\frac{\mathop{\pm} \ii n_\text{e} A_{\ell m}-\mu_{1,2} B_{\ell m}}{\kappa_{\mathrm{r}1,\mathrm{r}2}\sqrt{2\varepsilon_0}}\,,\\
    \Tilde{H}_{\ell m}^{(\pm1)}&=\frac{\mathop{\pm} \ii n_\text{e} B_{\ell m}+\varepsilon_{1,2} A_{\ell m}}{\kappa_{\mathrm{r}1,\mathrm{r}2}\sqrt{2\mu_0}}\,,
    \end{aligned}
\end{equation}%
where $n_\text{e}=k_z/k_0$ is the effective refractive index of the mode. Notice that since we used Hankel functions (rather than Macdonald $\macdonald{\ell}$) this form is the same inside and outside the fibre as long as one uses the right material parameters for the region, including relative permittivity $\varepsilon_{1,2}$, relative permeability $\mu_{1,2}$ and relative radial wavenumber $\kappa_{\mathrm{r}1,\mathrm{r}2}=\kappa_{1,2}/k_0$ such that $\kappa_{\mathrm{r}1,\mathrm{r}2}^2=\varepsilon_{1,2} \mu_{1,2} - n_\text{e}^2$. Note the electric and magnetic amplitudes are related by the electromagnetic duality transformation 
\begin{equation*}
    (\varepsilon,\tilde{E}_{\ell m}^{(s)};A_{\ell m},B_{\ell m})\mapsto(\mu,\tilde{H}_{\ell m}^{(s)};B_{\ell m},-A_{\ell m})\,.
\end{equation*}

\section{Boundary conditions, dispersion relation and mode amplitudes}
The boundary conditions require that fields parallel to the interface (in this case, longitudinal and azimuthal) are continuous. The longitudinal components are already continuous thanks to \cref{eq:z_component}. The azimuthal components lead to the boundary conditions of the form 
% \paco{[This doesnt seem trivial at all, not worth an appendix? you could also "hide" \cref{eq:ne2relationship} in that appendix because it is not easy to interpret and not used for anything later]}:
\begin{equation}\label{eq:boundary}
\begin{pmatrix}
\ell n_\text{e}& \ii\Phi_\ell^\mu 
\\  -\ii\Phi_\ell^\varepsilon & \ell n_\text{e}
\end{pmatrix}
\!\!
\begin{pmatrix}
A_{\ell m} \\ B_{\ell m}
\end{pmatrix}\! = \!
\begin{pmatrix}
0 \\ 0
\end{pmatrix}
\!.
\end{equation}
with {$u=\kappa_1\rho_0=k_0 \rho_0(n_1^2-n_\mathrm{e}^2)^{1/2}$, $w=-\ii\kappa_2\rho_0 = -\ii k_0 \rho_0(n_2^2-n_\mathrm{e}^2)^{1/2}$} being real (in the lossless case) dimensionless parameters satisfying {$V^2=u^2+w^2= k_0^2 \rho_0^2 (n_1^2-n_2^2)$}, with $V$ being called the normalised frequency.
We have also defined an analytical function: 
\begin{equation}\label{eq:phi}
    \Phi_\ell^\alpha(n_\text{e},V)=\qty(\frac{uw}{V})^2\qty(\frac{\alpha_1}{u} \frac{\bessel{\ell}^{\prime}(u)}{\bessel{\ell}(u)} +  \frac{\alpha_2}{w}  \frac{\macdonald{\ell}^{\prime}(w)}{\macdonald{\ell}(w)}),
\end{equation}
which also depends on the core and cladding material parameters $\alpha_{1,2}\in\{\varepsilon_{1,2},\mu_{1,2}\}$. 

\begin{figure*}[ht!]
  \centering
  \includegraphics[width=\textwidth]{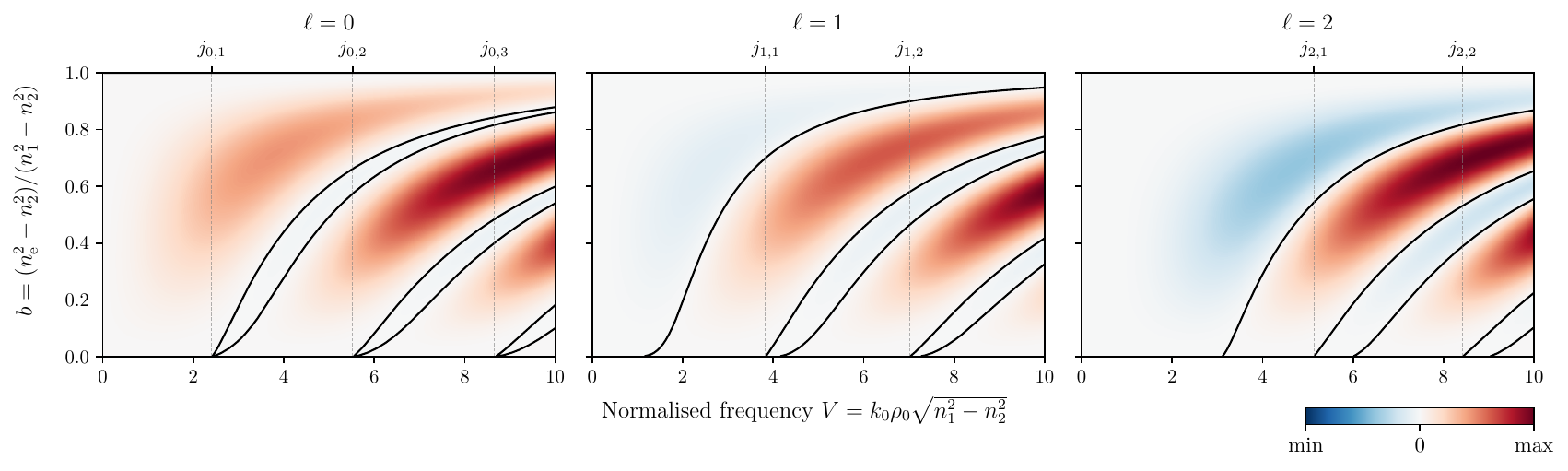}\vspace{-1em}
  \caption{The colour-map of the function $F_\ell(b,V)$ and its roots as solid black lines.}
  \label{fig:dispersion}
\end{figure*}

The dispersion relation comes from the condition under which \cref{eq:boundary} has non-trivial solutions, which is precisely when the determinant of the matrix is zero:\footnote{For $\ell\neq0$ \cref{eq:disp} reduces to
the relationship \(\varepsilon_\text{e}\mu_\text{e}=n_\text{e}^2\,,\)
with effective permittivity $\varepsilon_\text{e}=\Phi_\ell^\varepsilon/\ell$ and effective permeability $\mu_\text{e}=\Phi_\ell^\mu/\ell$. Interestingly, these effective material properties may be negative.}
\begin{equation}\label{eq:disp}
    {\Phi_\ell^\varepsilon\,\Phi_\ell^\mu}=({\ell}n_\text{e})^2\,,
\end{equation}
which is a transcendental equation with more than one solution (labelled by $m\in\{1,2,3,\ldots\}$) that has to be solved numerically.
This can be done by finding the roots of the function:
\begin{equation}\label{eq:dispersion_F}
    F_\ell(n_\text{e},V)=[{\Phi_\ell^\varepsilon\,\Phi_\ell^\mu}-({\ell}n_\text{e})^2]\bessel{\ell}^2(u),
\end{equation}
which is regular everywhere---the $\bessel{\ell}^2(u)$ is added to avoid singularities of \cref{eq:phi} at Bessel zeros---and which changes sign, making root finding very robust (see \cref{fig:dispersion}). The dispersion relations are zero contours of this function, $(n_\text{e})_{\ell m}(V)$, where for each azimuthal order $\ell$ the index $m$ labels the roots from high to low value of $n_\mathrm{e}$.
For implementation, it is more convenient to define $F_\ell(b,V)$ as a function of normalised propagation constant $b=(n_\mathrm{e}^2-n_2^2)/(n_1^2-n_2^2)\in[0,1]$, which can be understood as quantifying how strongly the mode is confined to the core of the fibre. The effective refractive index is then:
\begin{equation}
    n_\text{e}(b)=\sqrt{bn_1^2+(1-b)n_2^2}\,,
\end{equation}
while $w=V\sqrt{b}$, and $u=V\sqrt{1-b}$. The advantage is that the search interval is then a fixed range $[0,1]$ regardless of the materials of the core and fibre, and having found $b(V)$, it is straightforward to obtain $n_\text{e}$.

The amplitudes for a single eigenmode $\vc{F}_{\ell m}$ can be found from \cref{eq:boundary} to be a simple function of $n_\text{e}$:
\begin{equation}
\label{eq:analytical_AB}
\begin{pmatrix}
A_{\ell m} \\ B_{\ell m}
\end{pmatrix}
= \frac{N_{\ell m}}{\sqrt{1+|\nu_{\ell m}|^2}}
\begin{pmatrix}
1 \\ \nu_{\ell m}
\end{pmatrix}
,
\end{equation}
where $N_{\ell m}$ is an arbitrary constant and the only degree of freedom, which we may choose to be such that the field has unit power (see \cref{app:norm}),
and the parameter
\begin{equation}\label{eq:nu}
\nu_{\ell m}=\frac{\ii\Phi_{\ell m}^\varepsilon}{\ell n_\text{e}}=\frac{\ii \ell n_\text{e}}{\Phi_{\ell m}^\mu}\,, \!\!\quad
\nu_{0 m}=
    \begin{cases}
        0 &\text{if} \quad\Phi_{0m}^\varepsilon=0\\
        \ii\infty &\text{if}\quad \Phi_{0m}^\mu=0
    \end{cases},
\end{equation}
which represents the relationship between the longitudinal magnetic and electric fields,
is fixed by the dispersion relation, where the subscript $m$ in $\Phi_{\ell m}^\alpha$ emphasises that it is evaluated for a particular solution to \cref{eq:disp}. The use of \cref{eq:analytical_AB} to find the mode amplitudes analytically from $n_\text{e}$ instead of requiring a numerical null space finding procedure is a key advantage of this method.

\section{Naming of the modes and degeneracies}
In the scalar case where $\ell=0$, the transcendental \cref{eq:disp} can be separated into two distinct equations:
\begin{equation*}
    \!\!{\Phi_{0n}^\mu}=0=E_z \;(\text{TE}_{0n}),  \!\quad\text{or}\quad\!\Phi_{0n}^\varepsilon=0=H_z \;(\text{TM}_{0n})
    \,,\!\!
\end{equation*}
and depending on which is satisfied, the mode is either transverse electric or transverse magnetic. The $n\in\{1,2,3,\ldots\}$ is the radial order, which not only labels the roots but is also related to the number of radial nodes of the longitudinal part $n-1$.
Since these modes will have no $\varphi$ dependency, the fields have to be symmetric under rotations along $z$. As a consequence, the $\vc{E}$ (or $\vc{H}$) field in the TE (or TM) mode has to be azimuthally polarised, while the $\vc{H}$ (or $\vc{E}$) field will be transverse-spinning (see \cref{app:ellzero}).
These modes will be non-degenerate (they will have different $n_\text{e}$), unless $\Phi_0^\varepsilon=\Phi_0^\mu$, which would happen only if both the core and cladding are made of a material with $\varepsilon_\text{r}=\mu_\text{r}$. 

When $\ell\neq0$, the fields vary with $\varphi$ and cannot be purely TE or TM. Looking at \cref{eq:disp}, we can notice that the dispersion relation is invariant under the sign change $\ell\mapsto-\ell$. From that, we can expect that the eigenmodes $\vc{F}_{\pm|\ell|m}$ will be degenerate. These modes have a right(left)-handed quasi-circular polarisation respectively, but since they are degenerate, we can linearly combine them to get any polarisation:
\begin{equation}
    \vc{F}^{(\textsc{hem})}_{\ell m}=a_{+|\ell|m}\vc{F}_{|\ell|m}+a_{-|\ell|m}\vc{F}_{-|\ell|m}\,,
\end{equation}
which spans a Poincaré sphere with Stokes parameters:
\begin{alignat*}{4}
\mathcal{S}_0 &=|a_+|^2+|a_-|^2,%\\
&\quad
\mathcal{S}_2  &=2\Im(a_+a_-^*)\,, \\
\mathcal{S}_1&=2\Re(a_+a_-^*)\,,%\\
&\quad
\mathcal{S}_3  &=|a_+|^2-|a_-|^2\,,
\end{alignat*}
such that if $\mathcal{S}_3=0$ we have a quasi-linear polarisation. These combinations are called hybrid modes and should be labelled $\text{HEM}_{\ell m}$ according to IEEE standards (formerly IRE) \cite{ire1953_antennas_waveguides}. Notice that if $\varepsilon_r=\mu_r$ for both the core and cladding, TE$_{0n}$ and TM$_{0n}$ are degenerate,
and it is possible to linearly combine them into HEM$_{0n}$ modes
(hybrid because their linear combination will necessarily have $E_z$ and $H_z$ nonzero).
In the same dual-symmetric case, one also has $|E_z|=\eta_0|H_z|$ for all HEM$_{\ell m}$ with $\ell\neq0$. But outside of such a special scenario, it is common to adopt an alternative naming scheme based on longitudinal components:
\begin{equation*}
    \!\!|E_z|>\eta_0|H_z| \;(\text{HE}_{\ell n}),  \!\quad\text{or}\quad\!|E_z|<\eta_0|H_z| \;(\text{EH}_{\ell n})\,,\!\!
\end{equation*}
where the radial order is now given by $n=\frac{m}{2}+\frac{1-(-1)^m}{4}$ such that the $m$ index in $\text{HEM}_{\ell m}$ keeps alternating between $\text{HE}_{\ell n}$ ($\text{HEM}_{\ell(2n-1)}$) and $\text{EH}_{\ell n}$ ($\text{HEM}_{\ell(2n)}$) modes. In general (for example, if $\mu_\text{r}\neq1$), the magnitude of longitudinal components can no longer be used to distinguish between the two mode families. {We believe} one should look instead at the relative phase between the $E_z$ and $H_z$ encoded by $\Im(E_zH_z^*)(\rho_0)=\Im(AB^*)=|A||B|\sin(\arg{A}-\arg{B})$:
\begin{equation*}
    \Im(AB^*)>0 \;(\text{HE}_{\ell n}),  \!\quad\text{or}\quad\!\Im(AB^*)<0 \;(\text{EH}_{\ell n})\,,\!\!
\end{equation*}
which is a more general definition of the two families that is directly related to the transverse field patterns \cref{eq:spin_amplitudes}.

\section{Summary of the calculation}
Given a cylindrical fibre with a core radius $\rho_0$ and material relative permittivity and permeability in the core  $(\varepsilon_{1},\mu_{1})$ and in the cladding $(\varepsilon_{2},\mu_{2})$, with corresponding refractive indices $n_1=\sqrt{\varepsilon_{1} \mu_{1}}$ and $n_2=\sqrt{\varepsilon_{2} \mu_{2}}$, we wish to find the dispersion relation of each supported mode. Modal electromagnetic fields have a phase $k_z z - \omega t$ evolving in space $z$ and time $t$ in accordance with the dispersion relation that links the propagation constant $k_z$ and the angular frequency $\omega$. These two variables can be expressed in dimensionless form as the effective index $n_\text{e} = k_z/k_0$, which is further normalised to the unit interval $b\in[0,1]$ via $b=(n_\mathrm{e}^2-n_2^2)/(n_1^2-n_2^2)$, and the normalised frequency $V = k_0 \rho_0 \sqrt{n_1^2-n_2^2}$, where $k_0=\omega/c_0=2\pi/\lambda_0$. Once a mode is identified by a specific pair $(b,V)$, equivalently $(k_z,\omega)$, we also aim to compute the electric and magnetic vector field distributions inside and outside the fibre. The steps below are implemented in the open-source Python package \href{https://github.com/KZL358/Polarisation-Singularities}{\textcolor{black}{\faGithub}} \cite{anafibre}, which reproduces the full workflow from root-finding to field reconstruction.

This section outlines how to implement the proposed method in a simple program in five simple steps:

\begin{enumerate}[leftmargin=*, label=\textbf{Step \arabic*}:, align=parleft, wide=5pt]
    \item Define parameters and functions:
    \begin{itemize}
        \item Specify the core radius $\rho_0$ and the core and cladding material parameters $(\varepsilon_\text{r},\mu_\text{r})$, which can optionally be functions of wavelength {(dispersive) and complex-valued (lossy)}.
        \item {Implement functions $\Phi_\ell^\alpha(b,V)$ for $\alpha\in\{\varepsilon,\mu\}$ in terms of the normalised propagation constant $b \in [0,1]$ using \cref{eq:phi} with $u=V\sqrt{1-b}$, $w=V\sqrt{b}$.} 
        \item Note that $f_\ell(x)=x{\macdonald{\ell}^{\prime}(x)}/{\macdonald{\ell}(x)}$ is a regular function of $x$, however, evaluating $\macdonald{\ell}(x)$ and its derivative first will lead to a numerical divergence at $x=0$.
    \end{itemize}

    \item Construct the dispersion function:
    \begin{itemize}
        \item Define $F_\ell(b,V)$ from \cref{eq:dispersion_F}, expressed in terms of the normalised propagation constant $b \in [0,1]$ {by recalling $n_\text{e}=\sqrt{b n_1^2+(1-b)n_2^2}$}.
    \end{itemize}

    \item Solve the dispersion relation:
    \begin{itemize}
        \item For each azimuthal order $\ell$ and chosen normalised frequency $V$, numerically find the roots of $F_\ell(b,V)$.
        \item For order $\ell=0$ it is better to solve $\Phi_{0n}^\mu(b,V)=0$ ($\Phi_{0n}^\varepsilon(b,V)=0$) for TE (TM) modes respectively.
        \item Each root corresponds to a guided mode and defines the effective index $n_\text{e}$, labelled by $m$. For $\ell\neq0$ odd modes $m=2n-1$ are HE and even modes $m=2n$ are EH.
    \end{itemize}

    \item Compute field amplitudes analytically:
    \begin{itemize}
        \item For each solution with $\ell\neq0$, calculate the ratio $\nu_{\ell m}$ from \cref{eq:nu}.
        \item {Optionally}, choose the normalisation constant $N_{\ell m}$ such that the total guided power is unity, $|N_{\ell m}|^2 = 1~\text{W}/(c_0\,\sigma_{\ell m})$, see \cref{eq:normalisation} in \cref{app:norm}. {If not interested in absolute field values, then $N_{\ell m}$ is a free parameter.}
        \item Obtain the mode amplitudes $(A_{\ell m},B_{\ell m})$ using the \cref{eq:analytical_AB}. {For solutions with $\ell=0$} one can use amplitudes $(A_{0 m},B_{0 m})=|N_{\ell m}|(0,\ii)$ and $(A_{0 m},B_{0 m})=|N_{\ell m}|(1,0)$ for TE and TM modes respectively or alternatively expressions in \cref{app:ellzero}.
        \item Reconstruct the full vectorial fields from \cref{eq:ansatz,eq:radial,eq:spin_amplitudes}. {In these equations, recall that $\vc{F}_{\ell m}$ stands for both electric $\vc{E}_{\ell m}$ and magnetic $\vc{H}_{\ell m}$ fields, $\kappa_{1,2}=k_0 (n_{1,2}^2-n_\mathrm{e}^2)^{1/2}$ and $\kappa_{\mathrm{r}1,\mathrm{r}2} = \kappa_{1,2}/k_0$.}
    \end{itemize}
\end{enumerate}

\section{Conclusions}

In this work, we have presented a robust and efficient semi-analytical method for calculating the guided electromagnetic modes of cylindrical step-index nanofibres. By exploiting the underlying symmetries of the problem and introducing a convenient normalisation of the field amplitudes, the standard textbook $4\times4$ boundary-matching system was shown to be analytically reducible to an equivalent $2\times2$ formulation. This reduction removes the need for numerical null space calculations and allows the modal amplitudes to be obtained analytically once the dispersion relation is solved.

The resulting dispersion equation is simple, well behaved, and particularly well suited for numerical root finding, while the analytical determination of the full vectorial fields ensures accurate evaluation of the longitudinal components that are crucial for chiral, vectorial, and near-field light–matter interactions. The method is general, applies to arbitrary material parameters (including dispersive and lossy media), and retains full compatibility with standard mode classifications.

Beyond its conceptual clarity, this reformulation provides a practical advantage for numerical implementations and device modelling, offering a reliable foundation for applications in nanophotonics, chiral quantum optics, and fibre-based sensing where precise control and characterisation of guided modes is essential.

\section*{Acknowledgments}
Authors acknowledge support from the EIC Pathfinder project CHIRALFORCE (Grant No.~101046961), funded by the Innovate UK Horizon Europe Guarantee (UKRI Project No.~10045438).

\appendix
\section{Simplified boundary conditions}\label{app:two-by-two}
The equation below is an intermediate form between \cref{eq:four-by-four} and \cref{eq:boundary}. 
Starting from \cref{eq:four-by-four}, the first two rows enforce $C=A$ and $D=B$, leaving two independent boundary conditions for $(A_{\ell m},B_{\ell m})$. 
Written in terms of $u=\kappa_1\rho_0$ and $w=-\ii\kappa_2\rho_0$, these conditions become:
\begin{equation}
\begin{pmatrix}
\ell n_\text{e}\qty(\frac{1}{u^2}+\frac{1}{w^2})& \frac{\ii {\mu_1}}{u}\frac{\bessel{\ell}'}{\bessel{\ell}}+ \frac{\ii {\mu_2}}{w}\frac{\macdonald{\ell}'}{\macdonald{\ell}} 
\\  \frac{{\varepsilon_1}}{\ii u}\frac{\bessel{\ell}'}{\bessel{\ell}}+ \frac{ {\varepsilon_2}}{\ii w}\frac{\macdonald{\ell}'}{\macdonald{\ell}} & \ell n_\text{e}\qty(\frac{1}{u^2}+\frac{1}{w^2})
\end{pmatrix}
\!\!
\begin{pmatrix}
A_{\ell m} \\ B_{\ell m}
\end{pmatrix}
\! = \!
\begin{pmatrix}
0 \\ 0
\end{pmatrix}
\!,
\end{equation}
from which \cref{eq:boundary} follows by dividing through by a factor $\frac{1}{u^2}+\frac{1}{w^2}=\frac{V^2}{u^2w^2}$ and introducing $\Phi_\ell^\varepsilon$ and $\Phi_\ell^\mu$.
\section{Determining spin components from Maxwell's equations}\label{app:max}
For the spin basis $\uv{e}_{\pm1}=(\uv{x}\pm\ii\uv{y})/\sqrt{2}$ we can define partial derivatives $\partial_{\pm1}=(\uv{e}_{\pm1}^*\vdot\grad)=(\partial_{x}\mp\ii\partial_{y})/\sqrt{2}$, then the curl can be written in this frame as follows:
\begin{equation}
    \begin{split}
    \uv{e}_{\pm1}^*\vdot(\curl\vc{F})&=\pm\ii\big[\partial_{\pm1}F^{(0)}-\partial_{z}F^{(\pm1)}\big]\,.
    \end{split}
\end{equation}
Maxwell's equations will be ($\vc{F}_\text{e}=\sqrt{\varepsilon_0}\vc{E}$, $\vc{F}_\text{m}=\sqrt{\mu_0}\vc{H}$):
\begin{equation}
    \curl\vc{F}_\text{e}=\ii k_0\mu_\text{r}\vc{F}_\text{m}\,,\quad
    \curl\vc{F}_\text{m}=-\ii k_0\varepsilon_\text{r}\vc{F}_\text{e}\,,
\end{equation}
and their longitudinal spin components will be:
\begin{equation}
    \begin{split}
        k_0\varepsilon_\text{r}{F}^{(\pm1)}_\text{e}&=\mp\big[\partial_{\pm1}F_\text{m}^{(0)}-\partial_{z}F_\text{m}^{(\pm1)}\big]\,,\\
         k_0\mu_\text{r}{F}^{(\pm1)}_\text{m}&=\pm\big[\partial_{\pm1}F_\text{e}^{(0)}-\partial_{z}F_\text{e}^{(\pm1)}\big]\,,\
    \end{split}
\end{equation}
and we can combine them such that everything depends only on the scalar components and use $\partial_zF=\ii k_z F$:
\begin{equation}
    \begin{split}
        (k^2-k_z^2){F}^{(\pm1)}_\text{e}&=\partial_{\pm1}(\mp k_0\mu_\text{r}F_\text{m}^{(0)}+\ii k_zF_\text{e}^{(0)})\,,\\
        (k^2-k_z^2){F}^{(\pm1)}_\text{m}&=\partial_{\pm1}(\pm k_0\varepsilon_\text{r}F_\text{e}^{(0)}+\ii k_zF_\text{m}^{(0)})\,,\
    \end{split}
\end{equation}
where operators $\partial_{\pm1}$ act as ladder operators for so-called spin-weighted cylindrical harmonics (Bessel/Hankel functions with a phase factor):
\begin{equation}
\begin{split}
    \partial_{\pm1}[\bessel{\ell}(\kappa\rho)\ee^{\ii \ell\varphi}]&=\pm\frac{\kappa}{\sqrt{2}}\bessel{\ell\mp1}(\kappa\rho)\ee^{\ii (\ell\mp1)\varphi}\,,\\
    \partial_{\pm1}[\hankel{\ell}(\kappa\rho)\ee^{\ii \ell\varphi}]&=\pm\frac{\kappa}{\sqrt{2}}\hankel{\ell\mp1}(\kappa\rho)\ee^{\ii (\ell\mp1)\varphi}\,,
\end{split}
\end{equation}
and same for the other functions, which leads to \cref{eq:spin_amplitudes}.

\section{TE/TM mode fields}\label{app:ellzero}
While \cref{eq:ansatz} is universal for every $\ell$, the scalar mode case with $\ell=0$ has extra symmetries. Notice that in that case, the only $\phi$ dependence is through:
\begin{equation}
    \ee^{\mp\ii \varphi}\uv{e}_{\pm1}=(\uv{\rho}\pm\ii\uv{\varphi})/\sqrt{2}\,,
\end{equation}
which means that in cylindrical coordinates, the fields have no azimuthal dependence. Not only that, but the TE (TM) mode will have $A=0$ ($B=0$) as well as the radial-variation part of the solution: 
\begin{equation}
    R_{s}(\rho)=
    \begin{cases}
        {\bessel{-s}(\kappa_1\rho)}/{\bessel{0}(\kappa_1\rho_0)}& \text{if} \quad\rho<\rho_0
        \\{\hankel{-s}(\kappa_2\rho)}/{\hankel{0}(\kappa_2\rho_0)}& \text{if} \quad\rho>\rho_0
    \end{cases},
\end{equation}
satisfying $R_{1}=-R_{-1}$ leading to the following TE fields:
\begin{equation}
\begin{split}
    \vc{E}^{(\textsc{te})}_{0 n}&=\qty(\frac{\mu_\text{r}R_{1}}{\kappa_\text{r}\sqrt{\varepsilon_0}}\uv{\varphi})|N_{0n}|\;\ee^{\ii k_z z-\ii \omega t}\,,\\
    \vc{H}^{(\textsc{te})}_{0 n}&=\qty(\frac{\ii R_{0}}{\sqrt{\mu_0}}\uv{z}-\frac{n_\text{e}R_{1}}{\kappa_\text{r}\sqrt{\mu_0}}\uv{\rho})|N_{0n}|\;\ee^{\ii k_z z-\ii \omega t}\,,
\end{split}
\end{equation}
where we assume $B=\ii |N_{0n}|$. The TM fields can be obtained similarly or via dual transformation, which is an equivalent of $A=-\ii|N_{0n}|$:
\begin{equation}
\begin{split}
    \vc{H}^{(\textsc{tm})}_{0 n}&=\qty(\frac{\varepsilon_\text{r}R_{1}}{\kappa_\text{r}\sqrt{\mu_0}}\uv{\varphi})|N_{0n}|\;\ee^{\ii k_z z-\ii \omega t}\,,\\
    \vc{E}^{(\textsc{tm})}_{0 n}&=\qty(\frac{n_\text{e} R_{1}}{\kappa_\text{r}\sqrt{\varepsilon_0}}\uv{\rho}-\frac{\ii R_{0}}{\sqrt{\varepsilon_0}}\uv{z})|N_{0n}|\;\ee^{\ii k_z z-\ii \omega t}\,.
\end{split}
\end{equation}
From this, it is clear that the $\vc{E}^{(\textsc{te})}_{0 n}$ and  $\vc{H}^{(\textsc{tm})}_{0 n}$ fields are azimuthally polarised, while the $\vc{H}^{(\textsc{te})}_{0 n}$ and  $\vc{E}^{(\textsc{tm})}_{0 n}$ have an elliptical polarisation with the planes of ellipses laying in the $\rho z$ plane, which gives them transverse spin.

\section{Power and normalisation}\label{app:norm}
The following power normalisation derivation works in the case of lossless material. It is sensible to normalise our eigenvectors $\vc{F}_{\ell m}$ such that the total power flow through the cross-section is unity. This power can be defined via the z-directed time-averaged Poynting vector, integrated over the cross-sectional area:
\begin{equation}
    P=\frac{1}{2}\int_0^{2\pi}\!\!\!\int_0^\infty[\Re(\vc{E}\cp\vc{H}^*)\vdot\uv{z}]\rho\dd{\rho}\dd{\varphi}\,.
\end{equation}
The integrand depends only on the transverse fields:
\begin{equation*}
\begin{split}
    % \varPi_z=
    \Re(\vc{E}\cp\vc{H}^*)\vdot\uv{z}&=\Re[{E}_x{H_y^*}-{E_y}{H_x^*}]\\&=\Im[{E}_+{H_+^*}-{E_-}{H_-^*}]\,,
\end{split}
\end{equation*}
where $F_\pm=\uv{e}_{\pm1}^*\vdot\vc{F}$ and we used $F_x=(F_++F_-)/\sqrt{2}$ and $F_y=\ii(F_+-F_-)/\sqrt{2}$. Since the phase factors of $E_\pm$ and $H_\pm$ are identical, the product ${E}_\pm{H_\pm^*}={E}^{(\pm1)}{H}^{(\pm1)*}$, which are precisely the components that are piecewise defined in \cref{eq:radial}.
After lengthy algebra, the power can be split into a sum of four integrals 
% \paco{[This is black magic. If this is a result of some relatively long or medium derivation, please state it "after some derivation" or something, because right now it sounds as a trivial step]} \bastien{I agree, those steps are indeed a lot of black magic and the way it was worded was not giving it a justice}:
% \begin{equation}
% \begin{split}
%     P=A{|N_{\ell m}|^2}\sum_{i=1}^2\bigg[&\frac{\varepsilon_i+\mu_i|\nu_{\ell m}|^2}{1+|\nu_{\ell m}|^2}\frac{k_zk_0}{{|\kappa_{i}|^2}}I^{(i)}_+\\
%     &-\frac{2\Im(\nu_{\ell m})}{1+|\nu_{\ell m}|^2}\frac{k_z^2+k_i^2}{{|\kappa_{i}|^2}}I^{(i)}_-\bigg]\,,
% \end{split}
% \end{equation}
% \begin{equation}
% \begin{split}
%     P=\frac{A{|N_{\ell m}|^2}}{1+|\nu_{\ell m}|^2}\sum_{i=1}^2&\bigg[{(\varepsilon_i+\mu_i|\nu_{\ell m}|^2})\frac{k_zk_0}{{|\kappa_{i}|^2}}I^{(i)}_+\\
%     &
%     -2\Im(\nu_{\ell m})\frac{k_z^2+k_i^2}{{|\kappa_{i}|^2}}I^{(i)}_-\bigg]\,,
% \end{split}
% \end{equation}
% \begin{equation}\label{eq:power}
% \begin{split}
%     \!\!P={c_0{|N_{\ell m}|^2}}\underbrace{\sum_{i=1}^2\pi\rho_0^2\qty(\frac{k_zk_0}{{|\kappa_{i}|^2}}\alpha_{\ell m}^+I^{(i)}_++\frac{k_z^2+k_i^2}{{|\kappa_{i}|^2}}\alpha_{\ell m}^-I^{(i)}_-)}_{\sigma_{\ell m}},\!\!
% \end{split}
% \end{equation}
\begin{equation}\label{eq:power}
\begin{split}
    \!\!P&={c_0{|N_{\ell m}|^2}}\sigma_{\ell m},\qq{with:}\\
    \sigma_{\ell m}&=
    \sum_{i=1}^2\pi\rho_0^2\qty(\frac{k_zk_0}{{|\kappa_{i}|^2}}\alpha_{\ell m}^+I^{(i)}_++\frac{k_z^2+k_i^2}{{|\kappa_{i}|^2}}\alpha_{\ell m}^-I^{(i)}_-)\,,\!\!
\end{split}
\end{equation}
\vspace{-.5em}%
where we have packaged the core radial integrals into 
\begin{equation}
    I_\pm^{(1)}=\int_0^{u}
    \frac{\bessel{\ell-1}^2(v)\pm\bessel{\ell+1}^2(v)}{2\bessel{\ell}^2(u)}
    \frac{v\dd{v}}{u^2}\,,
\end{equation}
and the cladding integrals into
\begin{equation}\label{eq:I2}
    I_\pm^{(2)}=\int^\infty_{w}
    \frac{\macdonald{\ell-1}^2(v)\pm\macdonald{\ell+1}^2(v)}{2\macdonald{\ell}^2(w)}
    \frac{v\dd{v}}{w^2}\,,
\end{equation}
and finally, the $\alpha$ coefficients are:
\begin{equation*}\vspace{.25em}
    \alpha_{\ell m}^+=\frac{\varepsilon_i+\mu_i|\nu_{\ell m}|^2}{1+|\nu_{\ell m}|^2}\,,\quad
    \alpha_{\ell m}^-=\frac{\Im(\nu_{\ell m}^*)}{1+|\nu_{\ell m}|^2}\,.
\end{equation*}
Despite their intimidating appearance, these integrals are analytical and do not require numerical integration. The core integrals have the following form:
\begin{equation*}
   I_+^{(1)}=\frac12-\dfrac{\ell^2}{2u^2}+\dfrac{\bessel{\ell}'(u)}{u\bessel{\ell}(u)}+\dfrac{\bessel{\ell}'^2(u)}{2\bessel{\ell}^2(u)},\quad
    I_-^{(1)}=\dfrac{\ell}{u^2}\,,
\end{equation*}
where, as in the main text, $u=\kappa_1\rho_0$. 
The cladding counterparts of these expressions can be written in terms of $ w=-\ii\kappa_2\rho_0$ as:
\begin{equation*}
    \begin{split}
        \!\!I_+^{(2)}=\frac12+\dfrac{\ell^2}{2w^2}-\dfrac{\macdonald{\ell}'(w)}{w\macdonald{\ell}(w)}-\dfrac{\macdonald{\ell}'^2(w)}{2\macdonald{\ell}^2(w)},\quad\!\!\!
        I_-^{(2)}=-\dfrac{\ell}{w^2}\,.\!\!
    \end{split}
\end{equation*}
Using these analytical expressions further reduces numerical noise, as we no longer need to perform numerical integration, and in particular, we don't need to worry about the infinite limit in \cref{eq:I2}.
Finally, if we choose the power carried by the mode to be exactly unity ($1\,\text{W}$), then the normalisation constant has to be:
\begin{equation}\label{eq:normalisation}
\begin{split}
    |N_{\ell m}|=\sqrt{\frac{1\,\text{W}}{c_0\,\sigma_{\ell m}}}\,,
\end{split}
\end{equation}
where $\sigma_{\ell m}$ is an effective cross section defined in \cref{eq:power}.
We can also further write $|\kappa_{1}|^2=u^2/\rho_0^2$ and $|\kappa_{2}|^2=w^2/\rho_0^2$ such that we can minimise the number of variables.
\section{Analytical Jacobians of the fields}\label{app:gradients}

Several near-field light--matter observables depend not only on the modal fields themselves but also on their spatial derivatives. 
An example of such observables is dipolar optical forces, which depend on field gradients (Jacobians) $\grad\otimes\vc{E}$ and $\grad\otimes\vc{H}$, as well as quantities such as canonical momenta \cite{Golat2024,golat2024electromagnetic}. 
Analytical expressions for the Jacobians are therefore useful both for physical interpretation and for robust numerical implementation, since they avoid finite-difference noise in regions where the fields vary rapidly (notably near the fibre surface).

Starting from the modal expansion in the spin basis \cref{eq:ansatz}, and noting that the spin basis vectors $\uv{e}_s$ are spatially constant (being defined in the Cartesian frame), the gradient acts only on the scalar factor
$F_{\ell m}^{(s)}(\rho)\ee^{\ii(\ell-s)\varphi}\ee^{\ii k_z z}$.
Writing the gradient in cylindrical coordinates $\grad=\uv{\rho}\,\partial_\rho+\uv{\varphi}\,\rho^{-1}\partial_\varphi+\uv{z}\,\partial_z$ one finds the compact outer-product form:
\begin{equation*}
    \grad\otimes\vc{F}_{\ell m}=\ee^{-\ii \omega t}\sum_{s=-1}^1 (\ii\vc{k}_{\ell m s})\otimes\uv{e}_{s}\;F_{\ell m}^{(s)}\ee^{\ii (\ell-s)\varphi}\ee^{\ii k_z z}
    \,,
\end{equation*}
where $\vc{k}_{\ell m s}$ is an effective local wavevector for each spin component,
\begin{equation*}
    \vc{k}_{\ell m s}=({k_z})_{\ell m}\uv{z}+\frac{\ell-s}{\rho}\uv{\varphi}-\ii\kappa_{1,2}\frac{\harmol{\ell-s}'(\rho)}{\harmol{\ell-s}(\rho)}\uv{\rho}\,.
\end{equation*}
Here $\harmol{\nu}$ denotes the radial cylindrical harmonic used in the given region (e.g.\ $\harmol{\nu}(\rho)=\bessel{\nu}(\kappa_1\rho)$ in the core and $\harmol{\nu}(\rho)=\hankel[(1)]{\nu}(\kappa_2\rho)$ in the cladding; equivalently, one may use $\macdonald{\nu}(-\ii\kappa_2\rho)$ for guided lossless cladding), and $\kappa_{1,2}$ is the corresponding radial wavenumber.

\hbadness 10000\relax
\bibliography{main}

\end{document}